\documentclass{osa-article}

\journal{oe}

\articletype{Research Article}

\usepackage{amsmath,amsfonts,amssymb}
\usepackage{mathtools} 

\usepackage{bm}
\usepackage{comment}

\renewcommand{\sp}{\hspace{1pt}}
\newcommand{\nsp}{\hspace{-1pt}}

\newcommand{\ds}{\displaystyle}

\newcommand{\ud}{\mathrm{d}}
\newcommand{\bu}{\mathbf{u}}
\newcommand{\bv}{\mathbf{v}}
\newcommand{\bx}{\mathbf{x}}
\newcommand{\be}{\mathbf{e}}
\newcommand{\bq}{\mathbf{q}}

\newcommand{\bs}{\mathbf{s}}
\newcommand{\bE}{\mathbf{E}}
\newcommand{\lrangle}[1]{\langle #1 \rangle}

\newcommand{\NA}{\mathrm{NA}}

\renewcommand{\Re}{\mathrm{Re}}
\renewcommand{\Im}{\mathrm{Im}}

\usepackage{upgreek}

\newcommand{\JonesMat}[5][0]{\left[\hspace{-2pt}
	\renewcommand\arraystretch{.77}
	\begin{array}{cc}
		#2 & #3 \\[#1 pt]
		#4 & #5
	\end{array}
	\hspace{-2pt}
	\right]
}

\usepackage{scalefnt}
\makeatletter
\newcommand\mfrac[2]{%
	\dfrac{\text{\raisebox{-0.5ex}{\scalefont{0.85}{$\m@th#1$}}}}%
	{\text{\raisebox{0.35ex}{\scalefont{0.85}{$\m@th#2$}}}}%
}
\makeatother

\newcommand{\tablefrac}[2]{#1/#2} 

\DeclarePairedDelimiter{\norm}{\lVert}{\rVert}
\DeclarePairedDelimiter{\abs}{\lvert}{\rvert}

\usepackage{placeins}

\DeclareFontFamily{U}{BOONDOX-calo}{\skewchar\font=45 }
\DeclareFontShape{U}{BOONDOX-calo}{m}{n}{
  <-> s*[1.05] BOONDOX-r-calo}{}
\DeclareFontShape{U}{BOONDOX-calo}{b}{n}{
  <-> s*[1.05] BOONDOX-b-calo}{}
\DeclareMathAlphabet{\mathcalboondox}{U}{BOONDOX-calo}{m}{n}
\SetMathAlphabet{\mathcalboondox}{bold}{U}{BOONDOX-calo}{b}{n}
\DeclareMathAlphabet{\mathbcalboondox}{U}{BOONDOX-calo}{b}{n}

\newcommand{\mathscr}[1]{\mathcalboondox{#1}}

\usepackage{array,booktabs,multirow} 
\newcolumntype{M}[1]{>{\centering\arraybackslash}m{#1}}
\usepackage{longtable,tabularx}
\newcolumntype{x}[1]{>{\centering\arraybackslash\hspace{0pt}}p{#1}} 

\usepackage{xcolor}
\definecolor{dkgreen}{rgb}{0,.5,0}

\usepackage{soul}

\usepackage{makecell} 

\begin{document}

\title{Optimal birefringence distributions for star test polarimetry}

\author{Anthony Vella\authormark{1} and Miguel A.~Alonso\authormark{1,2,*}}
\address{\authormark{1}The Institute of Optics, University of Rochester, Rochester NY 14627, USA\\
   \authormark{2}Aix Marseille Univ, CNRS, Centrale Marseille, Institut Fresnel, UMR 7249, 13397 Marseille Cedex 20, France
	}
\email{\authormark{*}miguel.alonso@fresnel.fr}



\begin{abstract}
Star test polarimetry is an imaging polarimetry technique in which an element with spatially-varying birefringence is placed in the pupil plane to encode polarization information into the point-spread function (PSF) of an imaging system. In this work, a variational calculation is performed  to find the optimal birefringence distribution that effectively encodes polarization information while producing the smallest possible PSF, thus maximizing the resolution for imaging polarimetry. This optimal solution is found to be nearly equivalent to the birefringence distribution that results from a glass window being subjected to three uniformly spaced stress points at its edges, which has been used in previous star test polarimetry setups.
\end{abstract}


\section{Introduction}
Polarimetry is the measurement of the polarization state of light and/or the polarization properties of materials. Such measurements are usually characterized in terms of the Stokes parameters and the Mueller matrix, respectively, which are directly accessible from measurements of the intensity. Imaging polarimetry, in which the polarization is measured as a function of position, is particularly important in applications ranging from microscopy to remote sensing. 

Conventional techniques for Stokes polarimetry require multiple intensity measurements, either through time-sequencing or by splitting different polarization components into several separate detection channels \cite{Lacasse_2011,Azzam_2016}. For example, one common method uses a rotating quarter-wave plate (QWP) followed by a fixed linear polarizer, in which the Stokes parameters are deduced from successive intensity measurements with the QWP oriented at different angles \cite{Collett_1993}. 
While these techniques can produce highly accurate measurements, they can be relatively complicated and/or time-consuming, generally involving moving parts or multiple beam paths. 

When a short acquisition time is desirable, rapid polarization measurements may be taken using \emph{single-shot polarimetry}, in which the Stokes parameters are estimated from a single intensity measurement. A variety of methods exist for single-shot polarimetry, involving gradient index lenses \cite{Chang_2014}, patterned nanoscale gratings \cite{Chang_2014}, or a split aperture composed of multiple polarizers \cite{Chipman_1995_handbook}. One particularly simple method, referred to as \emph{star test polarimetry}, uses a spatially-varying birefringent mask (BM) followed by a uniform polarization analyzer in the pupil plane of an exit-telecentric imaging system. With an appropriately chosen birefringence distribution, the inserted elements can encode full polarization information into the shape of the PSF in the rear focal plane of the lens. This method has been demonstrated experimentally using a stress-engineered optic (SEO), which is a BK7 glass window subjected to stress with trigonal symmetry at its periphery, followed by a circular analyzer \cite{Ramkhalawon_2012, Ramkhalawon_2013}. Natural applications for this approach are those in which the object is a sparse set of discrete points, such as in astronomy and confocal microscopy. A recent specific application was the real-time monitoring of the output polarization states of each core within a multicore fiber bundle for applications in medical endoscopes  \cite{Sivankutty_2016}. 
An extension of this technique is being implemented within the context of superresolution microscopy for determining the 3D position, orientation, and vibration of independent fluorescent molecules \cite{Curcio_2019}. 
Star test polarimetry can also be applied for imaging continuous objects if these are discretized by placing a pinhole array in the object plane of an afocal $4f$ relay system with unit magnification \cite{Zimmerman_2014,Zimmerman_2016}. In all these applications, the polarization state of each object point can then be deduced from the corresponding PSF in the image plane, provided that the PSFs from different points do not overlap significantly. The spatial resolution of the measurement is thus limited by the size of the PSF. Encoding polarization information in the PSF necessarily implies increasing its size, since more information is being included in it. 


The goal of this work is to find the birefringence distribution of the mask in the pupil plane that maximizes spatial resolution by producing the smallest possible PSF while imposing reasonable restrictions to ensure that the object's polarization is effectively encoded in the shape of the PSF. The optimal solution is found to give very similar results to those of the birefringence distribution of the SEO. The solution's statistical performance (i.e., the expected error in the retrieved Stokes parameters) is assessed by calculating the Fisher information matrix for the measurement.

\section{System layout and notation}\label{sect:layout}
\begin{figure}
	\centering
	\includegraphics[scale=1]{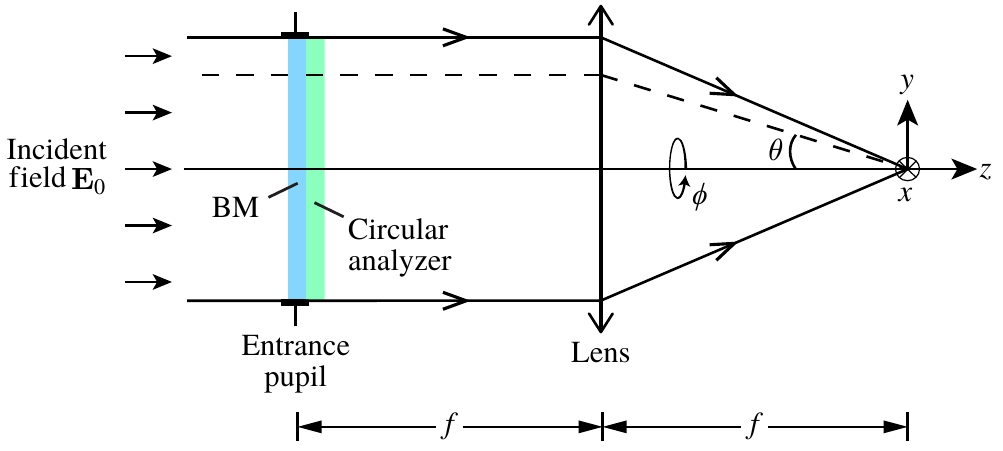}
	\caption{System layout for a star test polarimetry measurement of an unknown input field $\bE_0$, illustrated for the case in which a single circular polarization component is imaged. The pupil plane is described by a radial coordinate $u=\sin\theta$ and an azimuthal angle $\phi$.}
	\label{fig:system_layout}
\end{figure}

Consider the system shown in Fig.~\ref{fig:system_layout}, in which a spatially-varying, transparent, thin birefringent mask with Jones matrix $\mathbb{J}(\bu)$ is placed at the pupil plane of an exit-telecentric imaging system, where $\bu=(u\cos\phi,u\sin\phi)$ is the pupil coordinate normalized such that $u\leq\NA$, with NA being the system's numerical aperture. 
Collimated light (e.g., from a distant localized source) with uniform polarization $\bE_0$ is then incident on this BM in the plane of an aperture with pupil function $A(\bu)$ (binary or apodized) and is focused by a lens, producing a polarization-dependent PSF. The input polarization may then be deduced from the 
shape of the PSF by separating two orthogonal polarization components of the field, denoted by $\be_1$ and $\be_2$, before forming an image of one (or both) of them. For applications where photons are not scarce, only one of the two images is often necessary, so the extraction of the component in question (say, $\be_1$) can be performed with a polarizer. As will be discussed later, however,  there are applications in which it is best to use all photons, so instead the two components ($\be_1$ and $\be_2$) are separated by using a polarizing beamsplitter or a Nomarski or Wollaston prism and then they are imaged separately \cite{Curcio_2019}. 

In what follows, all calculations are performed in the circular polarization basis, so that $\be_1=(1,0)$ and $\be_2=(0,1)$, corresponding to right-hand circular (RHC) and left-hand circular (LHC) polarizations, respectively. If other output polarization components were analyzed, the results would still be valid after minor modifications. It is assumed that the imaging lens is slow (typically around 0.05 NA) so that the PSFs are imaged onto several pixels. All calculations  then assume the paraxial limit, and angle-dependent polarization effects at the lens can be neglected.

\section{Birefringence distribution and point-spread function}\label{sect:PSF}
We use here the notation introduced in Ref.~\cite{Vella_2018_poincare} to describe spatially-varying birefringence, where the Jones matrix of the BM in the circular polarization basis may be written as 
\begin{equation}
\mathbb{J}(\bu)
=e^{i\Gamma(\bu)}\!\JonesMat[2]{q_0(\bu)+i q_3(\bu)\;}{-q_2(\bu)+i q_1(\bu)}{q_2(\bu)+i q_1(\bu)\;}{\phantom{-}q_0(\bu)-i q_3(\bu)}\nsp,\label{eq:Jones}
\end{equation}
with $\Gamma(\bu)$ being a global phase function, and
\begin{subequations}\label{eq:qn}
	\begin{align}
	q_0(\bu) &= \cos\delta(\bu),\\
	q_1(\bu) &= \sin\delta(\bu)\cos\Theta(\bu)\cos\Phi(\bu),\\
	q_2(\bu) &= \sin\delta(\bu)\cos\Theta(\bu)\sin\Phi(\bu),\\
	q_3(\bu) &= \sin\delta(\bu)\sin\Theta(\bu),
	\end{align}
\end{subequations}
where $\Theta(\bu)\in[-\pi/2,\pi/2]$ and $\Phi(\bu)\in[0,2\pi)$ are the latitude and longitude on the Poincar\'e sphere of the ``slow'' eigenpolarization at the pupil point $\bu$, and $\delta(\bu)$ represents half the retardance between the two eigenpolarizations. As discussed in Ref.~\cite{Vella_2018_poincare}, the four quantities in Eq.~(\ref{eq:qn}) can be used to construct a four-dimensional unit vector $\vec{q}\sp(\bu)=(q_0,q_1,q_2,q_3)$, and $\mathbb{J}(\bu)$ itself can be considered as a unit quaternion. A particularly simple visualization of the birefringence results from using the 3-vector $\bq=(q_1,q_2,q_3)$, which points in the direction joining the two eigenpolarizations in the Poincar\'e sphere, and whose magnitude encodes the half-retardance $\delta$ as $|\bq(\bu)|=\sin\delta(\bu)$ \cite{Vella_2018_poincare}. Because $\vec{q}$ and $-\vec{q}$ correspond to the same effective birefringence, this vector can always be chosen such that $q_0=(1-|\bq|^2)^{1/2}\ge0$.

After transmission through the BM, the $\be_1$ and $\be_2$ polarization components of the field are given by $\be_1^\dagger\sp \mathbb{J}(\bu)\sp\bE_0$ and $\be_2^\dagger\sp \mathbb{J}(\bu)\sp\bE_0$, respectively, where the dagger denotes a conjugate transpose. The paraxial PSF $I^{(j)}(\bx)$ of each polarization component $\be_j$ is the squared modulus of the Fourier transform of the pupil distribution:
\begin{equation}
I^{(1,2)}(\bx) = \left\langle\,
\abs*{\iint A(\bu)\sp \be_{1,2}^\dagger\sp \mathbb{J}(\bu)\sp\bE_0 \exp[ik(\bu\cdot\bx)]\sp \ud^2u
}^2
\,\right\rangle_\mathrm{T},\label{eq:PSF_def}
\end{equation}
where $k=2\pi/\lambda$ is the wavenumber, $\bx$ is the two-dimensional spatial coordinate in the image plane, and $\langle\cdot\rangle_\mathrm{T}$ denotes a time average in the case of partially polarized light. As shown in Ref.~\cite{Vella_2018_poincare}, the phase function $\Gamma(\bu)$ causes the PSF to increase in size without helping to encode polarization information. Therefore, its optimal value is a constant, assumed here without loss of generality to be zero.

The PSF may be rewritten succinctly by introducing the pupil functions
\begin{equation}
g(\bu)=A(\bu)\sp[q_0(\bu) + i q_3(\bu)], \qquad	h(\bu)=A(\bu)\sp[-q_2(\bu) + i q_1(\bu)] \label{eq:gh}
\end{equation}
and their Fourier transforms
\begin{equation}
G(\bx)=\nsp\iint g(\bu)\exp[ik(\bu\cdot\bx)]\ud^2u,\qquad 
H(\bx)=\nsp\iint h(\bu)\exp[ik(\bu\cdot\bx)]\ud^2u. \label{eq:GH}
\end{equation}
Note that $G(\bx)$ and $H(\bx)$ represent the output $\be_1$ field component that results from an RHC or LHC polarized incident beam, respectively. As shown in  Appendix \ref{sect:app_PSF}, the PSF of each polarization component may then be expressed in terms of the Stokes parameters $S_0,S_1,S_2,S_3$ of the incident field $\bE_0$ in the form
\begin{equation}
I^{(1,2)}(\bx)=\mfrac{1}{2}\sum_{n=0}^3 S_n\mathcal{I}_n^{(1,2)}(\bx),\label{eq:I12_stokes}
\end{equation}
where the normalized intensity contributions $\mathcal{I}^{(1,2)}_n(\bx)$ are given by
\begin{subequations}\label{eq:In}
	\begin{alignat}{2}
	\mathcal{I}_0^{(1,2)}(\bx) &= |G(\pm\bx)|^2+|H(\pm\bx)|^2,\label{eq:I_0}\\
	\mathcal{I}_1^{(1,2)}(\bx) &= \pm 2\sp\Re\{G^*(\pm\bx)H(\pm\bx)\},\label{eq:I_1}\\
	\mathcal{I}_2^{(1,2)}(\bx) &= \pm 2\sp\Im\{G^*(\pm\bx)H(\pm\bx)\},\label{eq:I_2}\\
	\mathcal{I}_3^{(1,2)}(\bx) &= \pm \bigl(|G(\pm\bx)|^2-|H(\pm\bx)|^2\bigr).\label{eq:I_3}
	\end{alignat}
\end{subequations}
As shown in Appendix~\ref{sect:app_Fisher}, the Fisher information matrix for a measurement of the normalized Stokes vector $\bs=(S_1,S_2,S_3)/S_0$ when $\mathcal{N}$ photons are detected is given by 
\begin{equation}
[\sp\mathcal{N}\sp\mathbb{F}(\bs)]_{mn} = \frac{\mathcal{N}}{1+\overline{\bm{\mu}}\cdot\bs} 
\left[\,\overline{\left(\frac{\mu_m\mu_n}{1+\bm{\mu}\cdot\bs}\right)} - \frac{\overline{\mu_m}\;\overline{\mu_n}}{1+\overline{\bm{\mu}}\cdot\bs} \right]\nsp,
\label{eq:Fishergen}
\end{equation}
where, in what follows, we assume that only the image for the first polarization component ($\be_1$) is being used, so that $\bm{\mu}(\bx)$ is defined by
\begin{equation}
\bm{\mu}(\bx)=\frac{1}{\mathcal{I}_0^{(1)}(\bx)}
  \!\left[\!\nsp\renewcommand\arraystretch{.8}
  \begin{array}{c}
  \mathcal{I}_1^{(1)}(\bx) \\ \mathcal{I}_2^{(1)}(\bx) \\ \mathcal{I}_3^{(1)}(\bx)
  \end{array}
  \!\nsp\right]\nsp
\label{eq:mu}
\end{equation}
and $\overline{f}=\iint\! f(\bx)w(\bx)\ud^2 x$ represents a weighted integral of a function $f$ with weight 
$w(\bx)=\mathcal{I}_0^{(1)}(\bx)/\iint\! \mathcal{I}_0^{(1)}(\bx)\ud^2 x$. 
Note that $\bm{\mu}(\bx)$ is a unit vector, and therefore $|\overline{\bm{\mu}}|\le1$. The case in which both images are used is discussed in Section~\ref{section:bothPSFs}.

\section{Constraints on the BM distribution}\label{sect:constraints}
As mentioned previously, the primary performance metric for a candidate BM distribution is its impact on the width of the resulting PSF. However, it is necessary to impose some constraints in order to avoid solutions that would be unsuitable for polarimetry. For practical purposes, it is useful for the PSFs corresponding to each polarization component to have the same total power $\Psi^{(j)}=\iint\! I^{(j)}(\bx)\ud^2 x$. This makes the signal-to-noise ratio roughly independent of the input polarization, and it ensures that all polarization information encoded by the BM is fully contained within the shape of each intensity distribution rather than the overall power. Consequently, the input polarization can be deduced by analyzing the PSF of a single component, which is guaranteed to contain roughly half of the incident photons when the signal is sufficiently large. 

The total power of each polarization component can be found by integrating Eq.~(\ref{eq:I12_stokes}) over $\bx$ and applying Parseval's theorem to each term, leading to
\begin{align}
\Psi^{(1,2)}
	&= 	\mfrac{1}{2} \!\iint \Bigl[
		S_0\bigl(\abs{g}^2+\abs{h}^2\bigr) 
		\pm 2 S_1\Re\!\left\{g^*h\right\}
		\pm 2 S_2\Im\!\left\{g^*h\right\}
		\pm S_3\bigl(\abs{g}^2-\abs{h}^2\bigr)\Bigr]\ud^2u \nonumber\\
	&= \mfrac{1}{2} S_0 \!\iint\!\! A^2\,\bigl( 1 \pm \bm{\beta}\cdot\bs\bigr)\ud^2 u,
\end{align}
where $\bm{\beta}(\bu)=(-2q_0q_2+2q_1q_3,\,2q_0q_1+2q_2q_3,\,q_0^2-q_1^2-q_2^2+q_3^2)$ is a unit vector.
Then $\Psi^{(1)}=\Psi^{(2)}=\frac{1}{2}S_0\int\! A^2\ud^2 u$ for any arbitrary input polarization if $\langle\bm{\beta}\rangle_A\equiv\iint A^2(\bu)\bm{\beta}(\bu)\ud^2u=\mathbf{0}$. Note that this condition is not satisfied when $\vec{q}$ is constant, ruling out the possibility of a uniform BM.
In fact, from Parseval's theorem it is easy to see that $\overline{\bm{\mu}}=\langle\bm{\beta}\rangle_A/\Psi^{(1)}$.  Therefore, the condition $\langle\bm{\beta}\rangle_A=\mathbf{0}$ implies that $\overline{\bm{\mu}}=\mathbf{0}$, which causes a significant simplification of the Fisher information matrix in Eq.~(\ref{eq:Fishergen}),
in particular making zero the second term inside the brackets that is guaranteed to be non-positive for the diagonal elements of the matrix.  

\section{Derivation of differential equation and boundary conditions}\label{sect:optimization}
When the BM is introduced into the system, the PSF must increase in size since it now contains polarization information. Because the PSF would provide information about four values (the Stokes parameters) rather than one, the width of $I^{(j)}(\bx)$ can be expected to be roughly twice as large as that of a system with zero (or uniform) birefringence. Of course, there are many measures for the size of the PSF, each of which would give a slightly different result. The metric chosen here is the RMS irradiance width, which allows a simple variational calculation. It is expected that the optimal birefringence distribution for this measure will be nearly optimal for other measures, such as the full width at half maximum (FWHM) or the Strehl ratio.

Generally speaking, the width of the PSF does not strongly depend on the incident polarization state since the normalized intensity contributions $\mathcal{I}^{(1)}_n(\bx)$\;($n=0,1,2,3$) have similar compositions. Therefore, for simplicity, the performance may be evaluated in terms of the spread of the average PSF over all possible input polarizations, i.e., over all values of $S_1$, $S_2$, and $S_3$ within the range $[-S_0,S_0]$. This average produces $I^{(1)}(\bx)=\frac{1}{2}S_0\mathcal{I}_0^{(1)}(\bx)$, which is the PSF for unpolarized light. (Since $\mathcal{I}_0^{(1)}(\bx)=\mathcal{I}_0^{(2)}(-\bx)$, the analysis is also valid when the PSFs of both polarization components are used.) As shown in Ref.~\cite{Vella_2018_poincare}, the squared RMS width $r^2$ of the PSF for unpolarized light is
\begin{equation}
r^2=\frac{\iint \abs{\bx}^2 I^{(1)}(\bx)\,\ud^2x}{\iint\! I^{(1)}(\bx)\,\ud^2x}
=\frac{1}{\kappa}\iint \!\left(\abs{\nabla A}^2 + A^2\norm{\nabla\vec{q}\,}^2 \right)\ud^2u,
\label{eq:rsquared}
\end{equation}
where $\nabla$ is the gradient with respect to $\bu$, $\norm{\nabla\vec{q}}$ denotes the Frobenius norm of the $2\times4$ matrix $\nabla\vec{q}$, and $\kappa=k^2\iint\! A^2\ud^2u$. (This result can also be derived from Eqs.~(\ref{eq:I12_stokes}) and (\ref{eq:I_0}).) The term involving $\abs{\nabla\nsp A}^2$ accounts for diffraction, while the second term is the increase in width due to the birefringence distribution of the BM. Notice that if $A$ represents a hard aperture, the RMS width is not well-defined since $\nabla A$ diverges at the edge of the pupil. However, the increment
\begin{equation}
\Delta r^2 = 
\frac{1}{\kappa}\iint \! A^2\norm{\nabla\vec{q}\,}^2 \ud^2u
=\frac{1}{\kappa}\sum_{n=0}^3\iint \! A^2\nabla q_n\nsp\cdot\nsp\nabla q_n\, \ud^2u
\label{eq:Delta_r}
\end{equation}
caused by the BM can be well-defined and finite even for a hard-aperture pupil; it is this increment that we seek to minimize. Even for hard apertures, this measure of increase of the PSF width is expected to produce meaningful results, which can be verified by evaluating the FWHM of the solution for comparison.

Since $\vec{q}$ is a unit vector, it contains only three free components for optimization. A natural choice is to use the three-dimensional vector $\bq$ and let $q_0=(1-\abs{\bq}^2)^{1/2}$. 
To find the solutions we use a variational approach: we consider a linear combination of the PSF's width increase and the three constraints, $M=\kappa\,\Delta r^2+{\bm \Lambda}\cdot\langle{\bm \beta}\rangle_A$, where ${\bm\Lambda}=(\Lambda_1,\Lambda_2,\Lambda_3)$ is a vector of Lagrange multipliers and the constant prefactor $\kappa$ was introduced for future convenience. The variational procedure consists of finding the conditions under which $M$ is stationary under infinitesimal changes in $\bq$.
For this purpose, consider a variation $\vec{q}\to\vec{q}+\vec{\epsilon}$, where $\vec{\epsilon}=(\epsilon_0,\epsilon_1,\epsilon_2,\epsilon_3)$ is an infinitesimal change. Since after this change the vector must remain a unit vector, $\vec{\epsilon}$ must be perpendicular to $\vec{q}$, so only three of its components are independent parameters, e.g., the zeroth component can be chosen to be a function of the remaining ones according to 
\begin{equation}
\epsilon_0=-\frac{\bm{\epsilon}\cdot \bq}{q_0}. 
\label{eq:eps0}
\end{equation}

Let us first analyze the variations of each of the different parts, starting with the PSF width increase. It can be seen from Eq.~(\ref{eq:Delta_r}) that $\vec{q}\to\vec{q}+\vec{\epsilon}$ causes the change
\begin{align}
\Delta r^2 &\to \Delta r^2+
\frac{2}{\kappa}
	\sum_{n=0}^3\iint \! A^2\nabla q_n\cdot\nabla\epsilon_n\,\ud^2u\nonumber\\
	&=\Delta r^2-\frac{2}{\kappa}\sum_{n=0}^3\iint \! \epsilon_n\nabla\nsp\cdot\left(A^2\nabla q_n\right) \ud^2u+\left[\frac{2}{\kappa}\int_{\rm edge} \!A^2\vec{\epsilon}\cdot\partial_\perp\vec{q}\,\ud u\right],	
\label{eq:chDeltar1}
\end{align}
where we ignored terms quadratic in $\vec{\epsilon}$ and in the second step we removed the derivatives from $\epsilon_n$ by using integration by parts. The term in square brackets is an edge contribution that is present in the case of hard apertures, where $\partial_\perp$ denotes the derivative normal to the edge, and it should be included only if edges are excluded from the region of integration within the second term (since otherwise this edge contribution is already included in this integral). The corresponding variations in the three constrained quantities $\lrangle{\bm{\beta}}_A$ 
give
\begin{subequations}\label{eq:chbeta}
	\begin{align}
	\lrangle{\beta_1}_A &\to \lrangle{\beta_1}_A+2\textstyle\iint\! A^2(-\epsilon_0q_2+\epsilon_1q_3-\epsilon_2q_0+\epsilon_3q_1)\sp\ud^2u,\label{eq:chbeta1}\\
	\lrangle{\beta_2}_A &\to \lrangle{\beta_2}_A+2\textstyle\iint\! A^2(\epsilon_0q_1+\epsilon_1q_0
	+\epsilon_2q_3+\epsilon_3q_2)\sp\ud^2u,\label{eq:chbeta2}\\
	\lrangle{\beta_3}_A &\to \lrangle{\beta_3}_A+2\textstyle\iint\! A^2(\epsilon_0q_0-\epsilon_1q_1-\epsilon_2q_2+\epsilon_3q_3)\sp\ud^2u.\label{eq:chbeta3}
	\end{align}
\end{subequations}

In Eqs.~(\ref{eq:chDeltar1}) and (\ref{eq:chbeta}), $\epsilon_0$ can be expressed in terms of the remaining $\epsilon_n$ by using Eq.~(\ref{eq:eps0}). The substitution of the resulting variations into $M$ causes a variation of the form $M\to M+\iint(M_1\epsilon_1+M_2\epsilon_2+M_3\epsilon_3)\,\ud^2u$.  
The total variation of $M$ is then insensitive to the infinitesimal changes $\epsilon_n$ at any point inside the pupil only if $M_1=M_2=M_3=0$ everywhere inside the pupil, that is, if the following three constraints hold:
\begin{subequations}\label{eq:conds}
	\begin{align}
	q_1\nabla\!\cdot\nsp(A^2\nabla q_0)-q_0\nabla\!\cdot\nsp(A^2\nabla q_1)
	+A^2\nsp\bigl[\Lambda_1(q_0q_3+q_1q_2)+\Lambda_2(q_0^2-q_1^2)-2\Lambda_3q_0q_1\bigr]
	&=0,\label{eq:conds1}\\[2pt]
	q_2\nabla\!\cdot\nsp(A^2\nabla q_0)-q_0\nabla\!\cdot\nsp(A^2\nabla q_2)
	+A^2\nsp\bigl[\Lambda_1(q_2^2-q_0^2)+\Lambda_2(q_0q_3-q_1q_2)-2\Lambda_3q_0q_2\bigr]&=0,\label{eq:conds2}\\[2pt]
	q_3\nabla\!\cdot\nsp(A^2\nabla q_0)-q_0\nabla\!\cdot\nsp(A^2\nabla q_3)
	+A^2\nsp\bigl[\Lambda_1(q_0q_1+q_2q_3)+\Lambda_2(q_0q_2-q_1q_3)\bigr]&=0.\label{eq:conds3}
	\end{align}
\end{subequations}
Note that for hard apertures, these equations are dominated at the edges by the derivatives acting on $A^2$, which is discontinuous. However, as mentioned earlier, it is perhaps more convenient to account for variations of $M$ due to the edges by using the edge term in Eq.~(\ref{eq:chDeltar1}). After following similar steps, it is easy to find three edge constraints that can be written succinctly as
\begin{equation}
\label{eq:edge}
    q_0\,\partial_\perp\bq\,\Big|_{\rm edge}=\bq\,\partial_\perp q_0\,\Big|_{\rm edge}.
\end{equation}
This way,  Eqs.~(\ref{eq:conds}) apply within the smooth regions of the aperture (even infinitesimally close to the edges) and constitute the differential equations for $\bq(\bu)$ to be solved, while Eq.~(\ref{eq:edge}) applies at the edges, providing appropriate boundary conditions.

\section{Solution ignoring the boundary conditions}
In order to solve these equations, it is illustrative to first solve the simpler problem in which the constraints are ignored in the variational calculation (that is, the Lagrange multipliers are set to zero), as are the boundary conditions. In this case, Eqs.~(\ref{eq:conds1}) through (\ref{eq:conds3}) can be written compactly as 
\begin{equation}
q_0\sp\nabla\nsp\cdot\nsp(A^2\nabla\bq)=\bq\sp\nabla\nsp\cdot\nsp(A^2\nabla q_0)
\label{eq:Euler-Lagrange_q_eval_vec}
\end{equation}
or as a set of several equalities:
\begin{equation}\label{eq:Euler-Lagrange_q_eval_4}
\frac{\nabla\nsp\cdot\nsp(A^2\nabla q_0)}{q_0} 
  = \frac{\nabla\nsp\cdot\nsp(A^2\nabla q_1)}{q_1}
  = \frac{\nabla\nsp\cdot\nsp(A^2\nabla q_2)}{q_2}
  = \frac{\nabla\nsp\cdot\nsp(A^2\nabla q_3)}{q_3}.
\end{equation}
The latter representation highlights the underlying symmetry between the coordinates of $\vec{q}\sp(\bu)$ on the Poincar\'e hypersphere they inhabit~\cite{Vella_2018_poincare}.

Consider the standard case of a circular hard aperture with pupil function
\begin{equation}
A(\bu) = \begin{cases}1,&u\leq \NA,\\0& \text{otherwise}.\end{cases}
\end{equation}
The rotational symmetry of the pupil combined with the symmetries of the differential equations mentioned earlier allow solving the problem through separation of the polar pupil variables $u,\phi$. The solution becomes particularly simple if we assume that the BM is not birefringent at the pupil's center, i.e., that $\bq={\bf 0}$ for $u=0$. (Note that this choice causes no loss in generality, since the cascading of the resulting BM with plates with uniform birefringence before and/or after it gives access to other solutions.) It is easy to show that a set of solutions to Eq.~(\ref{eq:Euler-Lagrange_q_eval_vec}) is given by
\begin{equation}\label{eq:solsgeneral}
\bq(u,\phi)=\frac{2b_mu^{|m|}}{1+b_m^2u^{2|m|}}\left[\cos(m\phi),\sin(m\phi),0\right],
\end{equation}
for integer $m$, and where $b_m$ is a constant. (A derivation of this result in terms of a stereographic projection of the hyperspherical coordinates $\vec{q}$ is given in Ref.~\cite{Vella_2018_thesis}.) 
Note that we also made the choice of setting $q_3=0$. Again, there is no loss of generality in this choice, because other solutions can be found through cascading with uniform birefringent plates. 

\begin{figure}
\centering
\includegraphics[scale=1]{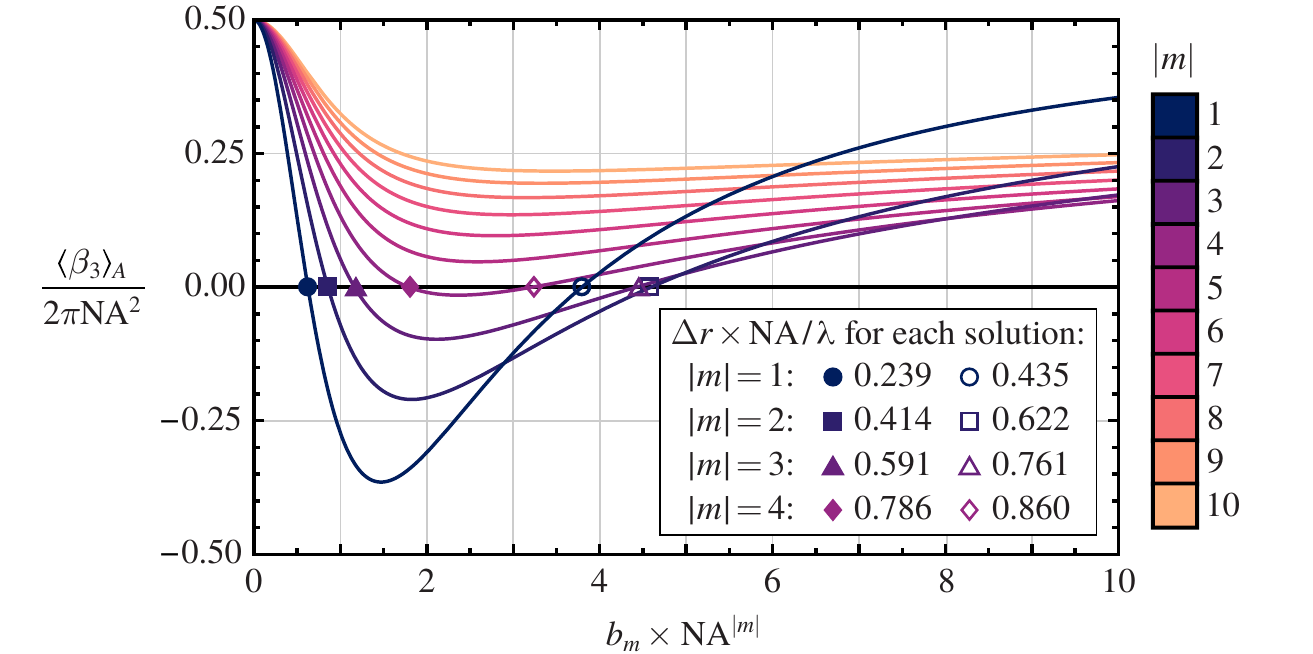}
\caption{Value of  Eq.~(\ref{eq:beta3_chi_0}) as a function of $b_m$ (times a scaling factor) for several values of $|m|$. Note that this quantity can vanish only for $1\leq|m|\leq 4$. The inset shows the corresponding values of $\Delta r$ for the eight values of $|m|$ and $b_m$ for which the condition $\langle\beta_3\rangle_A=0$ is satisfied.
}\label{fig:beta3_harmonics}
\end{figure}

The solutions in Eq.~(\ref{eq:solsgeneral}) turn out to automatically satisfy the constraints $\lrangle{\beta_1}_A=\lrangle{\beta_2}_A=0$. For the third constraint ($\lrangle{\beta_3}_A=0$), we calculate
\begin{align}
\frac{\langle\beta_3\rangle_A}{2\pi\NA^2} &= \frac1{\NA^2} \int_0^\NA  \frac{b_m^4u^{4|m|}-6\sp b_m^2u^{2|m|}+1}{(1+b_m^2u^{2|m|})^2} \sp u\sp\ud u\nonumber\\
&=\int_0^1  
	\frac{(b_m\NA^{|m|}v^{|m|})^4 - 6(b_m\NA^{|m|}v^{|m|})^2 + 1}{\bigl[1+(b_m\NA^{|m|}v^{|m|})^2\bigr]^2} \, v\sp\ud v,\label{eq:beta3_chi_0}
\end{align}
where we used the change of variables $v=u/\NA\in[0,1]$ in the last step. 
The value of this expression 
is plotted in Fig.~\ref{fig:beta3_harmonics} as a function of $b_m\NA^{|m|}$ for $|m|=1,2,\ldots,10$. 
Observe that when $1\leq|m|\leq 4$, there are exactly two values of $b_m$ for which $\langle\beta_3\rangle_A=0$. For each of these eight cases, the values of the increment in the RMS width of the PSF can be calculated by using
\begin{align}
\Delta r &= \frac{1}{2\pi} \frac{\lambda}{\NA}
  \left[\frac{1}{\pi}\sum_{n=0}^3 \int_0^{2\pi}\int_0^1 (\nabla_{\nsp\bv} q_n \cdot \nabla_{\nsp\bv} q_n) \, v \sp\ud v \ud\phi\right]^{1/2}\nonumber\\
  &= \frac{1}{2\pi} \frac{\lambda}{\NA}
  \left(2\int_0^1 \left\{\left[\bar{\delta}'(v)\right]^2+\frac{m^2}{v^2}\sin^2\bar{\delta}(v)\right\}\nsp v\sp \ud v\right)^{1/2}, \label{eq:delta_r_sol}
\end{align}
where $\bar{\delta}(v)=\delta(v\NA)$. Note from this expression that solutions with higher $|m|$ tend to cause larger increases in the size of the PSF. For the solutions in Eq.~(\ref{eq:solsgeneral}), $\bar{\delta}(v)=\frac12\arctan(b_m\NA^{|m|}v^{|m|})$. 
The numerical results obtained for each solution are shown in the inset in Fig.~\ref{fig:beta3_harmonics}, where we can see that the solution with $|m|=1$, $b_{1}=0.625/\NA$ produces the smallest increase in PSF width, which is $\Delta r=0.239\lambda/\NA$. The resulting half-retardance distribution is then $\bar{\delta}(v)= 2\arctan(0.625 v)$.

While the solutions found above satisfy the constraints $\langle\bm{\beta}\rangle_A=\bm{0}$, the fact that these constraints were ignored at the outset in the variational calculation means that they cannot simultaneously satisfy this constraint and the boundary condition in Eq.~(\ref{eq:edge}), which for the hard circular aperture considered here
(where the normal derivative $\partial_\perp$ reduces to a radial derivative) is given by
\begin{equation}
\bar{\delta}'(1)=0.
\label{eq:condsbord}
\end{equation}
These solutions are therefore not optimal, but they provide insights that are useful when considering the true optimal solution.

\section{Optimal birefringence distribution}
\label{section:optimal}
\begin{figure}
\centering
\includegraphics[scale=1]{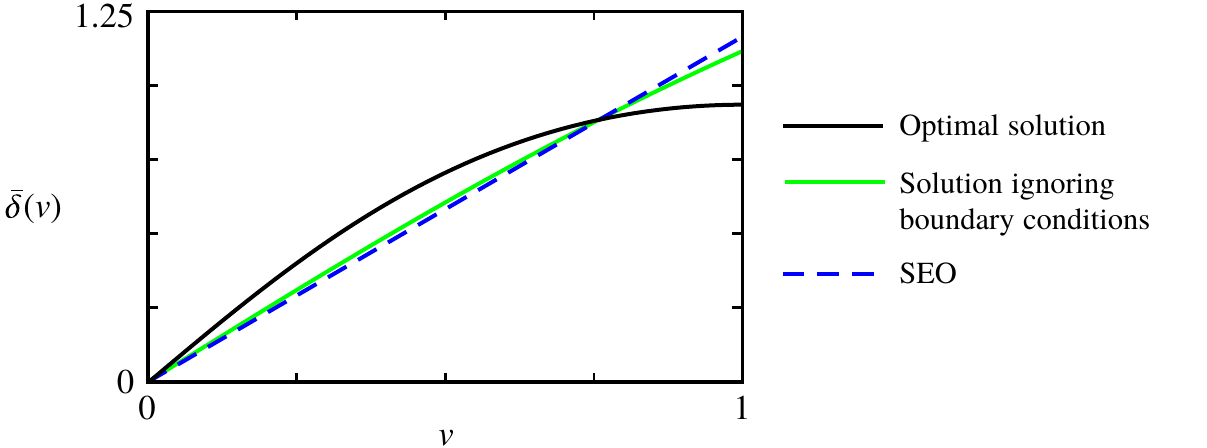}
\caption{Radial retardance distribution $\bar{\delta}(v)$ for the optimal BM solution ignoring the boundary conditions, the true optimum, and an SEO with stress coefficient $c=1.166/\NA$, plotted as functions of the normalized radial pupil coordinate $v$.}\label{fig:delta}
\end{figure}

Let us now go back to the general equations (\ref{eq:conds}) while again considering a hard circular pupil. Drawing inspiration from the previous case, we propose a separable solution constrained to the $q_1q_2$ plane, namely 
$\bq(u,\phi)=\sin\delta(u)\,[\cos(m\phi),\sin(m\phi),0]$ with $m\ne0$, which automatically satisfies the constraints $\lrangle{\beta_1}_A=\lrangle{\beta_2}_A=0$. Notice from Eqs.~(\ref{eq:qn}) that this ansatz corresponds to 
$q_0(u)=\cos\delta(u)$, $\Theta=0$, and $\Phi=m\phi$. 
The substitution of this ansatz into Eq.~(\ref{eq:conds3}) implies that 
$\Lambda_1=\Lambda_2=0$, and the resulting form of both  Eqs.~(\ref{eq:conds1}) and (\ref{eq:conds2}) gives a differential equation for the half-retardance $\delta$, which after changing variables to $v=u/\NA$ becomes
\begin{equation}
    \bar{\delta}''(v)+\frac{\bar{\delta}'(v)}v+\left(\NA^2\Lambda_3-\frac{m^2}{2v^2}\right)\sin\!\left[2\bar{\delta}(v)\right]=0,
\label{optdiffeq}
\end{equation}
for $v\in[0,1]$, where the assumption of no birefringence at the pupil center translates into the condition $\bar{\delta}(0)=0$, and the boundary condition in Eq.~(\ref{eq:condsbord}) must be satisfied. 
Additionally, the solution must satisfy the constraint $\langle\beta_3\rangle_A=0$, which implies $\int_0^1\cos\!\left[2\bar{\delta}(v)\right]\nsp v\sp\ud v=0$. Equation~(\ref{optdiffeq}) must be solved numerically subject to these boundary conditions and constraints by choosing appropriately the initial slope and $\NA^2\Lambda_3$.  Like for the case considered in the previous section, the optimal solution corresponds to $|m|=1$. This solution is shown in Fig.~\ref{fig:delta}, and it turns out to be very well approximated by the quadratic expression $1.856v-0.922v^2$ (the R-squared of the fit being 0.9996). This solution achieves an increase in the PSF width of $\Delta r=0.220\lambda/\NA$, which is indeed slightly smaller than that found in the previous section.

\section{Performance evaluation and comparison to SEO}\label{sect:variational_SEO}
As discussed in the introduction, star test polarimetry is implemented experimentally by using an SEO having three points of symmetric stress. 
The birefringence distribution near the center of an SEO turns out to be approximately given by the same general form as the previous two solutions, namely $\bq(u,\phi)=\sin\delta(u)\,[\cos(m\phi),\sin(m\phi),0]$, with $m=-1$ and where in this case the half-retardance is given by $\delta(u) = c\hspace{.5pt}u$,
with $c$ being a measure of the stress in the SEO \cite{Ramkhalawon_2012, Ramkhalawon_2013}.   
Given the similarity of this birefringence distribution to the optimal one, the constraints $\langle\beta_1\rangle_A=0$ and $\langle\beta_2\rangle_A=0$ are also automatically satisfied. The third constraint reduces to
\begin{equation}
\frac{\langle\beta_3\rangle_A}{2\pi\NA^2} = \frac{1}{\NA^2}\int_0^\NA \cos(2cu)u\sp\ud u 
= \left[\cos(c\NA)-\frac12\mathrm{sinc}(c\NA)\right]\!\mathrm{sinc}(c\NA)=0,
\end{equation}
where $\mathrm{sinc}(x)=x^{-1}\sin x$. 
This equation has  infinitely many solutions for $c$, but the one that leads to the smallest PSF width increase is the first one, $c=1.166/\NA$, shown in Fig.~\ref{fig:delta}.

Table~\ref{tbl:PSF_metrics} provides a comparison of the performance metrics of the BM solutions found in the previous two sections and the SEO. The 
half-retardance distributions are also summarized. All three solutions correspond to $|m|=1$. (The SEO corresponds specifically to $m=-1$, but the results are the same for $m=1$, which corresponds to an SEO followed a uniform half-wave plate.) The optimal solution does outperform the two other solutions not only in RMS width increase $\Delta r$ (calculated through Eq.~(\ref{eq:delta_r_sol})) for which it was optimized, but also for two other performance metrics: the FWHM and the Strehl ratio. However, the differences in these metrics between the optimal solution and the SEO are of only between 5 and 10\%. Also provided for comparison are corresponding metrics for a diffraction-limited imaging system where no BM is used (or equivalently where the BM is uniform). We can see that the cost of encoding polarization information into the PSF (on average over all possible polarizations) is an increase in its FWHM of about 60\% and a reduction of the Strehl ratio by about 50\%.
\begin{table}[tb]
\caption{Performance metrics for the PSFs resulting from using three BM distributions characterized by the half-retardance functions in the second column, as well as for a diffraction-limited system, when the incident field is unpolarized. 
}
\centering
{\small
\begin{tabular}{lll@{\hspace{20pt}}l@{\hspace{20pt}}l@{\hspace{20pt}}l}
\toprule
\phantom{a}\\[-20pt]
Solution & $\bar{\delta}(v)$ & $\Delta r$  & FWHM & Strehl ratio\\
\phantom{a}\\[-20pt]
\midrule 
Optimal 
  & $\approx\nsp 1.856v\nsp-\nsp 0.922v^2$
  & $\tablefrac{0.220\sp\lambda}{\NA}$ 
  & $\tablefrac{0.764\sp\lambda}{\NA}$ 
  & 0.490 \\
Solution w/o BC's
  & $\tfrac{1}{2}\!\arctan(0.625v)$
  & $\tablefrac{0.239\sp\lambda}{\NA}$ 
  & $\tablefrac{0.792\sp\lambda}{\NA}$ 
  & 0.474 \\
SEO 
  & $1.166v$
  & $\tablefrac{0.249\sp\lambda}{\NA}$ 
  & $\tablefrac{0.798\sp\lambda}{\NA}$ 
  & 0.469 \\
Diffraction limit
  & 0
  & {\small Not applicable}
  & $\tablefrac{0.514\sp\lambda}{\NA}$ 
  & 1.000 \\  
\bottomrule
\end{tabular}
}
\label{tbl:PSF_metrics}
\end{table}

\begin{figure}
\centering
\includegraphics[scale=1]{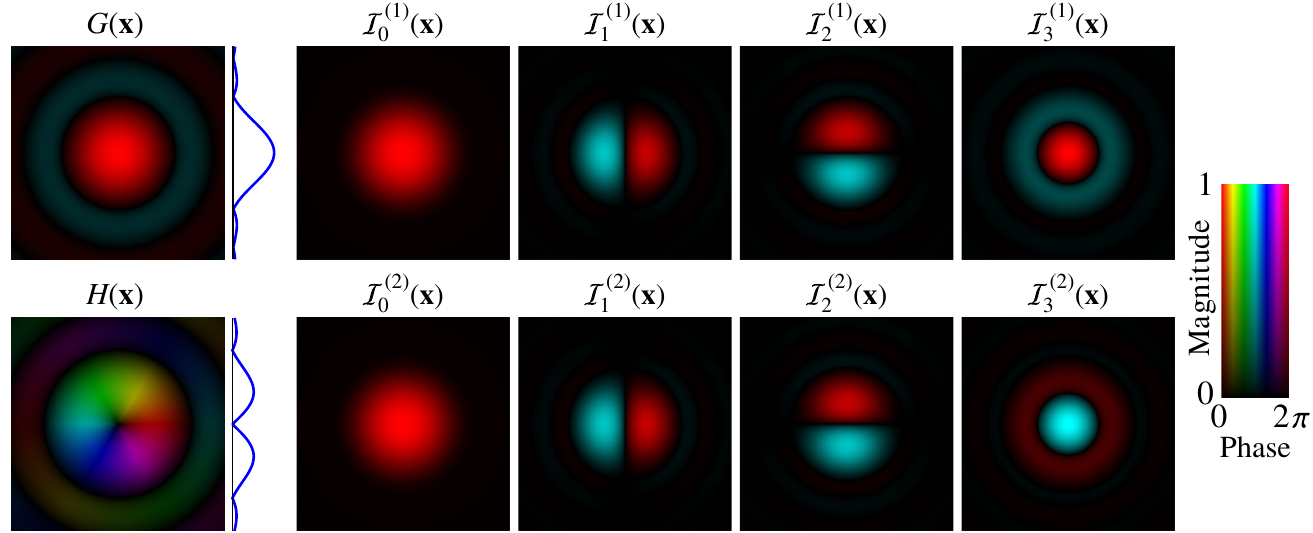}
\caption{For the optimal BM solution: complex fields $G$ and $H$ (left, with amplitude cross-sections through the center shown on the right), 
and PSF contributions $\mathcal{I}_n^{(j)}(\bx)$ 
to the PSFs of the $\be_1$ (top row) and $\be_2$ (bottom row) output polarization components. The plots are shown over a square region with half-width $1.25\lambda/\NA$.}
\label{fig:Icontr_opt}
\end{figure}

\begin{figure}
\centering
\includegraphics[scale=1]{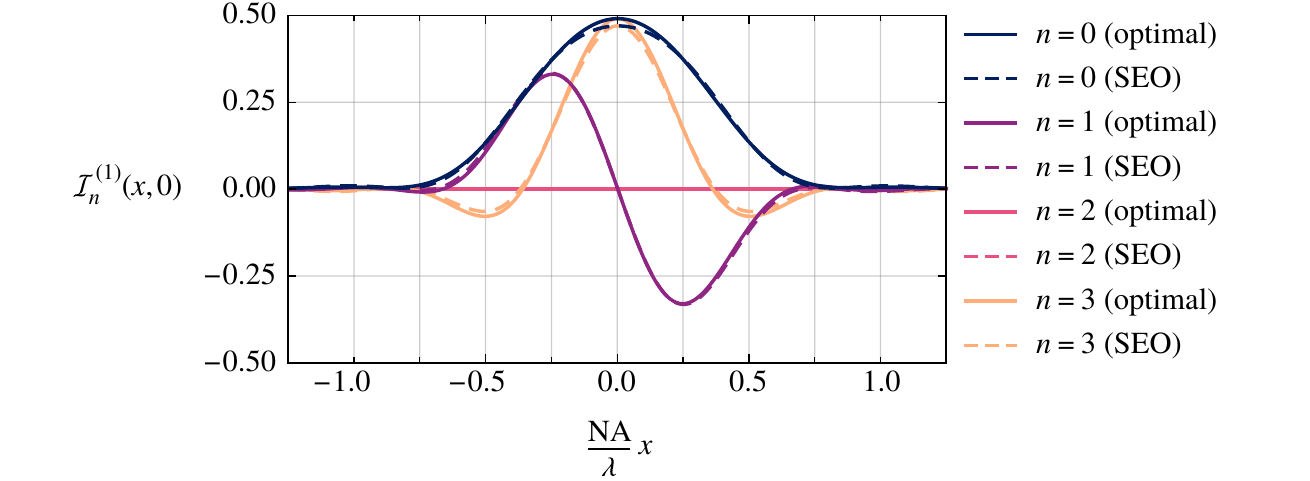}
\caption{Horizontal slices through $y=0$ of the intensity contributions $\mathcal{I}_n^{(1)}$. The solid and dashed curves correspond to the optimal birefringence distribution and an SEO, respectively.}\label{fig:Icontr_xsections}
\end{figure}

The functions $G(\bx)$ and $H(\bx)$ and the corresponding PSF contributions $\mathcal{I}_n^{(j)}(\bx)$ for the optimal BM are shown in 
Fig.~\ref{fig:Icontr_opt}. Note that the PSF contributions involve negative contributions for $n\ge1$, but the total measured PSF, which corresponds to the superposition in Eq.~(\ref{eq:I12_stokes}), is always non-negative as a result of the constraint $S_0^2\ge S_1^2+S_2^2+S_3^2$. For the PSF contributions, the top/bottom row corresponds to the case where the right/left circularly polarized component emerging from the BM is focused on the detector. Notice that, for these solutions, $\mathcal{I}_n^{(j)}(\bx)$ for $n=0,1,2$ are independent of $j$ (the circular component being used) while $\mathcal{I}_3^{(j)}(\bx)$ changes by a global sign. That is, the symmetries of the functions $G(\bx)$ and $H(\bx)$ are such that the only sign change ($\pm$) that has an effect in Eqs.~(\ref{eq:In}) is that in Eq.~(\ref{eq:I_3}). The equivalent plots for an SEO are virtually indistinguishable and are therefore not shown. Instead, to appreciate the very subtle differences, plots of the horizontal cross-sections $\mathcal{I}_n^{(1)}(x,0)$ for both the optimal BM solution and the SEO are shown in Fig.~\ref{fig:Icontr_xsections}. In practice, the effect of these differences is probably insignificant compared with variations that would arrive, for example, from the pixelization of the detector. 

\begin{figure}
\centering
\includegraphics[scale=1]{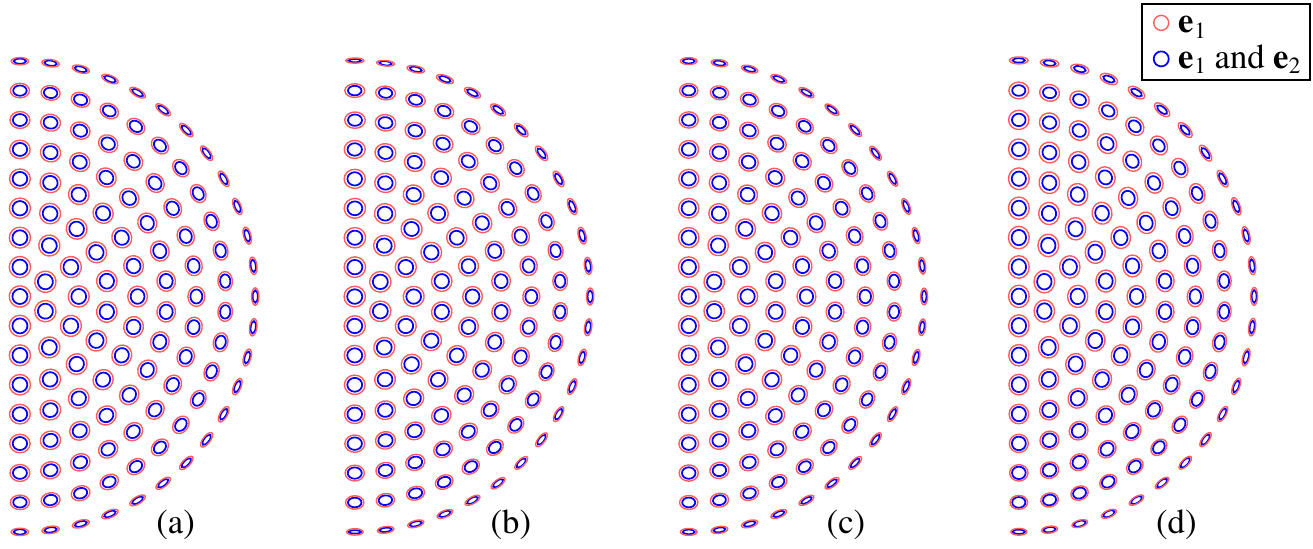}
\caption{Two-dimensional cross-sections of the error ellipsoids for incident polarization states in the $s_1$-$s_2$ and $s_1$-$s_3$ cross-sections (shown only for $s_1\ge0$ due to symmetry) of the Poincar\'e sphere for (a-b) the optimal BM and (c-d) an SEO, when only the $\be_1$ output polarization component is measured with $\mathcal{N}=1500$ photons being detected (red) and when both polarization components are measured with $\mathcal{N}=3000$ photons being detected (blue).}
\label{fig:ellipses}
\end{figure}

The accuracy of the measurement can be assessed by using the Fisher information matrix in Eq.~(\ref{eq:Fishergen}), which given the constraints imposed on the solutions reduces to
\begin{equation}
[\mathcal{N}\mathbb{F}(\bs)]_{mn}=\mathcal{N}\,\overline{\left(\frac{\mu_m\mu_n}{1+\bm{\mu}\cdot\bs}\right)}.
\label{eq:Fishershort}
\end{equation}
The inverse of this matrix provides an estimate of the variance in the accuracy of the measurements. That is, for a given incident polarization $\hat{\bs}$, the retrieved normalized Stokes parameter vector $\bs$ is expected (within one standard deviation) to be within an ellipsoid given by ${(\bs-\hat{\bs})\cdot[\mathcal{N}\mathbb{F}(\bs)]\cdot(\bs-\hat{\bs})=1}$.  Figure~\ref{fig:ellipses} shows cross-sections (in red) of these ellipsoids for several incident polarization states, both for the optimal BM and for the SEO, for the case of $\mathcal{N}=1500$ detected photons. The error ellipsoids are slightly more symmetric with respect to the $s_1$-$s_2$ plane for the optimal BM, but overall the performance is essentially the same. Note that these expected deviations of the normalized Stokes vectors scale as $\mathcal{N}^{-1/2}$, and that the width of the ellipsoids in the plane normal to the cross-sections is comparable to that in the azimuthal direction.

The fact that all solutions described so far have the form $\bq(u,\phi)=\sin\delta(u)\,[\cos(\phi),\pm\sin(\phi),0]$, where $\delta(u)$ is linear near the center, can be understood in terms of the two circular components of the incident light. 
Recall that the BM is followed by an RHC polarizer and a Fourier-transforming lens before reaching the detector. Since the birefringence is smallest at the center of the BM and grows away from it, the BM/RHC polarizer combination acts as a rotationally-symmetric apodizer for the incident RHC-polarized light. The incident LHC-polarized component, on the other hand, gets converted into RHC light away from the BM's center, and acquires a phase vortex due to a geometric phase effect. 
After Fourier transformation by the lens, the field distributions at the detector plane for these components are precisely the distributions $G$ and $H$ (shown in Fig.~\ref{fig:Icontr_xsections} for the optimal BM), whose forms are a localized central lobe with uniform phase and a slightly larger ``donut'' with a phase vortex of unit charge at its center, respectively. The radial extent of the latter would increase with $|m|$, the magnitude of the vortex charge written by the BM on 
$H$, 
so $|m|=1$ guarantees the most compact vortex distribution.
The superposition of these two fields then gives access to combinations with all relative phases between them, as well as with all relative amplitudes due to their different radial dependence.

The accuracy of a polarimetric measurement is known to increase with the volume within the Poincar\'e sphere enclosed by a simplex whose corners correspond to the polarization components being measured \cite{Tyo_2006}. The analysis above tells us then that, in theory, each point of the PSF being measured corresponds to the measurement of a different polarization component, and that these cover the complete surface of the Poincar\'e sphere in a fairly uniform way, so that all the interior of the sphere is enclosed. The uniformity of the coverage of the Poincar\'e sphere in the $s_3$ direction following from the variation in the radial direction of the relative weights of the two field components is examined in Fig.~\ref{fig:I3overI0}(a) for both the optimal solution and the SEO. While the uniformity is not perfect, all points are represented in similar amounts and the centroid of the distribution is zero, as guaranteed by the constraint $\overline{\mu_3}=0$. In practice, of course, the sampling of these different polarization components is compromised by detector pixelation, and even if the pixels were to be made very small, this would come at the cost of each detecting less photons. However, it is clear that this type of measurement provides a more ``democratic'' coverage of the sphere when compared, say, to a polarimeter where six polarization components are measured.

\section{Results when both polarization components are measured}
\label{section:bothPSFs}
So far we have assumed that only one circular polarization component ($\be_1$) emerging from the BM is focused to form an image. However, as mentioned in the introduction, it is possible to instead separately image both circular components and analyze the two resulting PSFs jointly. While doing so requires a more complex system, there are applications in which it is advantageous \cite{Curcio_2019}. One clear advantage is that the number of detected photons is essentially doubled, reducing the uncertainty in the estimation of the Stokes parameters by roughly a factor of $1/\sqrt{2}$. Note, however, that when both images are used, $\overline{\bm{\mu}}$ vanishes automatically (where the bar now indicates weighted integration over both images) because the total number of detected photons is independent of the polarization state. The Fisher information matrix is then given by the simpler form in Eq.~(\ref{eq:Fishershort}) without having to impose the constraints introduced in Section~\ref{sect:constraints}. While removing these constraints would clearly affect the variational derivation presented earlier, we now show that the results found earlier remain nearly optimal for the case when both components are imaged. 

Consider again the representative case of nearly unpolarized light ($\bs\approx{\bf 0}$), for which the uncertainty in the estimation of the Stokes parameters is largest, and for which the Fisher information matrix in Eq.~(\ref{eq:Fishershort}) reduces to $\mathcal{N}$ times $\overline{\mu_m\mu_n}$. Given the forms found for $\mathcal{I}_n^{(j)}(\bx)$ (antisymmetric in $x$ and symmetric in $y$ for $n=1$, symmetric in $x$ and antisymmetric in $y$ for $n=2$, and rotationally symmetric for $n=3$), the $3\times 3$ matrix with components $\overline{\mu_m\mu_n}$ is automatically diagonal. Further, because $\bm{\mu}(\bx)$ is a unit vector, the trace of this matrix is unity. The uncertainty in the measurement is then minimized if the three diagonal elements have equal magnitude, that is, if $\overline{\mu_n^2}=1/3$ for $n=1,2,3$. It turns out that the optimal BM solution derived in Section~\ref{section:optimal} nearly achieves this, giving $\overline{\mu_1^2}=\overline{\mu_2^2}=0.32$ and $\overline{\mu_3^2}=0.35$, while the SEO gives $\overline{\mu_1^2}=\overline{\mu_2^2}=0.35$ and $\overline{\mu_3^2}=0.30$. The cross-sections of the uncertainty ellipsoids for both the optimal BM solution and the SEO are shown (in blue) in Fig.~\ref{fig:ellipses} for the case in which $1500$ photons are detected in each PSF, giving a total number of photons of $\mathcal{N}=3000$. The first thing one notices is that the uncertainties are indeed smaller by a factor of about $1/\sqrt{2}$ with respect to those where only the PSF for the $\mathbf{e}_1$ component was used (red). Also, the uncertainties are now exactly symmetric in $s_3$, as is the coverage of $s_3$ shown in Fig.~\ref{fig:I3overI0}(b).

Finally, note that the relations $\overline{\mu_m\mu_n}=\overline{\mu_n^2}\,\delta_{mn}$ and $\overline{\bm{\mu}_n}={\bf 0}$, valid for the optimal and SEO solutions (and whether only $\be_1$ or both $\be_1$ and $\be_2$ components are imaged),  imply that the four basic PSFs, ${\cal I}_n^{(j)}(\bx)$, are orthogonal (and almost orthonormal since $\overline{\mu_n^2}\approx1/3$) under the weight $1/{\cal I}_0^{(j)}(\bx)$. This orthogonality is useful in the retrieval of the parameters from the measured PSFs.

\section{Concluding remarks}
\begin{figure}
\centering
\includegraphics[scale=1]{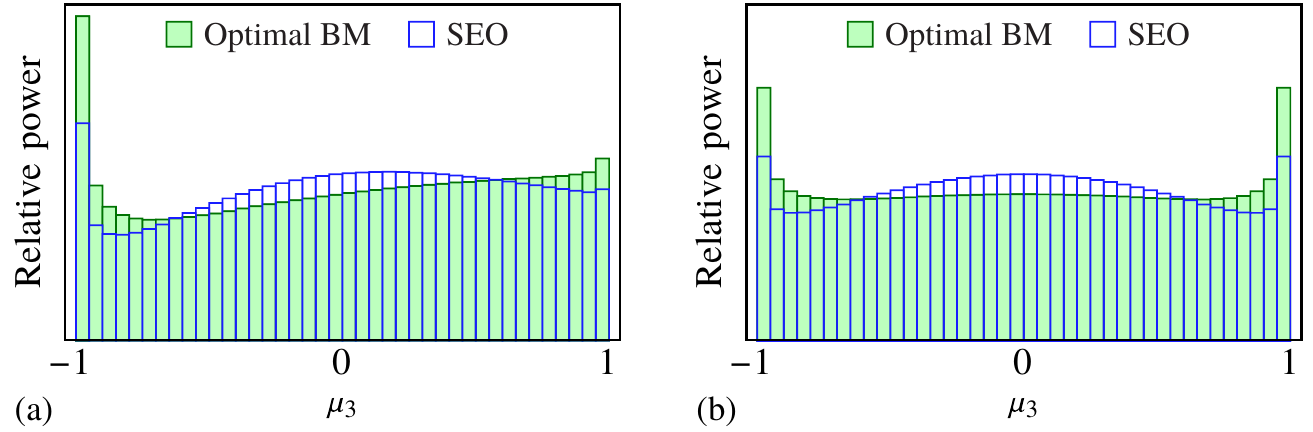}
\caption{Histograms of the power coverage of different values of $s_3$ (characterized by $\mu_3$) over the extent of the PSFs, for the cases when (a) only the PSF for the $\mathbf{e}_1$ component is used, and (b) when the PSFs for both components are used. }\label{fig:I3overI0}
\end{figure}

A variational calculation was performed to determine the spatial birefringence distribution to be placed at the pupil plane of an imaging system that optimizes the encoding of polarization in the PSF's shape while keeping the PSF size as small as possible. This optimal birefringence pattern was found to be similar (both in distribution and performance) to that found naturally near the equilibrium points of a transparent window under stress: the mask is not birefringent at its center, but its retardance grows with the radial pupil coordinate $u$, while the orientation of the slow and fast axes rotates with the azimuthal pupil coordinate $\phi$ as $\pm\phi/2$. (As mentioned earlier, other solutions with equal performance correspond to birefringence distributions that result from the cascading of these solutions with uniform wave plates.) The only small differences were in how the retardance grows with $u$ away from the center, but these cause only small variations in the resulting PSFs.  In fact, if the variational derivation had been based on a different measure of PSF size  and/or requirements for polarimetric optimality (e.g. a measure of the uniformity of $\mu_3$ in Fig.~\ref{fig:I3overI0}), the details of the optimal $\delta(u)$ would change slightly, these variations being comparable to those between any of these optimal solutions and the SEO. That is, the mechanics of stress-induced birefringence provide a simple way to produce a nearly optimal birefingence distribution for imaging polarimetry, without the need for nano-fabrication or combinations of spatial light modulators: by applying a pressure distribution with trigonal symmetry to the edges of a a glass window, a nearly-optimal birefringence mask (the SEO) results. 

Since the aim of this work was to show that the optimal BM distribution has very similar characteristics and performance as the SEO, we considered the continuous functional form of the PSFs. The performance of these devices in real applications, however, will also depend on the pixelation and general characteristics of the detector, as well as on the algorithms used for the retrieval of the Stokes parameters from the measured PSFs. Preliminary work on this direction can be found in Ref.~\cite{Vella_2018_thesis}. These aspects are particularly important in applications such as fluorescence microscopy \cite{Curcio_2019}, where photons are scarce and the PSFs are used to extract information not only about the two-dimensional polarization studied here, but of the three-dimensional polarization (as well as the three-dimensional position) of the light emitted by a fluorophore. For such applications, the near-optimality of the SEO is of central importance.

\section*{Appendix}
\appendix

\section{Point-spread function}\label{sect:app_PSF}
The polarization-dependent PSF from Section \ref{sect:PSF} 
can be derived as follows. Consider the pupil functions
\begin{equation}
	g_\pm(\bu)=A(\bu)\sp[q_0(\bu) \pm i q_3(\bu)],\qquad
	h_\pm(\bu)=A(\bu)\sp[]-q_2(\bu) \pm i q_1(\bu)]
\label{eq:gh_pm}
\end{equation}	
and their Fourier transforms
\begin{equation}
G_\pm(\bx)=\iint g_\pm(\bu)\exp[ik(\bu\cdot\bx)]\ud^2u,\qquad
H_\pm(\bx)=\iint h_\pm(\bu)\exp[ik(\bu\cdot\bx)]\ud^2u.
\label{eq:GH_pm}
\end{equation}
(Note that the ``$+$'' subscripts correspond to the quantities $g$, $h$, $G$, and $H$ defined in Eqs.~(\ref{eq:gh}) and (\ref{eq:GH}).) Then Eqs.~(\ref{eq:Jones}) and (\ref{eq:PSF_def}) lead to the following expression for the PSF:
\begin{equation}
I^{(1,2)}(\bx)=\left\langle\abs*{G_{\pm}(\bx)E_{1,2}\pm H_{\pm}(\bx)E_{2,1}}^2\right\rangle_\mathrm{T},
\end{equation}
where $E_1$ and $E_2$ are the right- and left-circular polarization components of the input field. This can be expanded to obtain
\begin{align}
I^{(1,2)}(\bx) &= \bigl\langle
			\abs{G_\pm}^2\abs{E_{1,2}}^2
			+ \abs{H_\pm}^2\abs{E_{2,1}}^2
			\pm 2\,\Re\{G_\pm^*H_\pm E_{1,2}^*E_{2,1}\}
			\bigr\rangle_\mathrm{T}\nonumber\\
		  &= \abs{G_\pm}^2\lrangle{\abs{E_{1,2}}^2}_\mathrm{T}
			+ \abs{H_\pm}^2\lrangle{\abs{E_{2,1}}^2}_\mathrm{T} \nonumber\\
		&\phantom{=}\quad \pm 2\,\Re\{G_\pm^*H_\pm\} \lrangle{\Re\{E_{1,2}^*E_{2,1}\}}_\mathrm{T}
			\mp 2\,\Im\{G_\pm^*H_\pm\} \lrangle{\Im\{E_{1,2}^*E_{2,1}\}}_\mathrm{T}\nonumber\\
		&= \mfrac{1}{2}\nsp\Bigl[
		   \abs{G_\pm}^2(S_0\pm S_3)
			+ \abs{H_\pm}^2(S_0\mp S_3)
			\pm 2\,\Re\{G_\pm^*H_\pm\} S_1
			+ 2\,\Im\{G_\pm^*H_\pm\} S_2 \Bigr],
\end{align}
where in the last step, the electric field was rewritten in terms of the Stokes parameters of the incident field, given by
\begin{alignat}{2}
S_0 &= \langle\abs{E_1}^2\rangle_\mathrm{T} + \langle\abs{E_2}^2\rangle_\mathrm{T},\qquad
&& S_1 = 2\langle \sp\Re\{ E_2^*E_1 \}\rangle_\mathrm{T},\quad \nonumber\\
S_2 &= 2\langle \sp\Im\{ E_2^*E_1 \}\rangle_\mathrm{T},\quad \,
&& S_3 = \langle\abs{E_1}^2\rangle_\mathrm{T} - \langle\abs{E_2}^2\rangle_\mathrm{T}. \label{eq:intro_Stokes}
\end{alignat}
Thus, the PSF can be written as 
\begin{equation}
I^{(1,2)}(\bx)=\mfrac{1}{2}\sum_{n=0}^3 S_n\mathcal{I}_n^{(1,2)}(\bx),\label{eq:app_polarimetry_I12_stokes}
\end{equation}
where the normalized intensity contributions associated with each Stokes parameter are given by
\begin{alignat}{2}
\mathcal{I}_0^{(1,2)}(\bx) &= |G_\pm(\bx)|^2+|H_\pm(\bx)|^2,\qquad
&&\mathcal{I}_1^{(1,2)}(\bx) = \pm 2\sp\Re\{G^*_\pm(\bx)H_\pm(\bx)\},\nonumber\\
\mathcal{I}_2^{(1,2)}(\bx) &= 2\sp\Im\{G^*_\pm(\bx)H_\pm(\bx)\},\qquad
&&\mathcal{I}_3^{(1,2)}(\bx) = \pm \bigl( |G_\pm(\bx)|^2-|H_\pm(\bx)|^2\bigr).
\label{eq:app_polarimetry_In_pm}
\end{alignat}
These results are simplified by noting that $g_-(\bu)=g_+^*(\bu)$ and $h_-(\bu)=h_+^*(\bu)$. From Eq.~(\ref{eq:GH_pm}), it follows that $G_-(\bx)=G_+^*(-\bx)$ and $H_-(\bx)=H_+^*(-\bx)$, leading to the form in Eq.~(\ref{eq:In}).

\section{Fisher information}\label{sect:app_Fisher}
The Fisher information matrix for a measurement of the normalized Stokes vector $\bs$ may be derived using the formalism shown in Ref.~\cite{Vella_MLE}. The probability density function for a measurement of the continuous PSF is defined as
\begin{equation}
P(\bx|\bs) = \frac{I^{(1)}(\bx)}{\iint\nsp I^{(1)}(\bx)\sp\ud^2 x}.
\end{equation}
Using Eq.~(\ref{eq:I12_stokes}), this expression can be written in terms of the normalized Stokes parameters as
\begin{align}
P(\bx|\bs) &= 
\frac{\ds\mathcal{I}_0^{(1)}\!\left(\! 1 + \sum_{n=1}^3 s_n \frac{\mathcal{I}_n^{(1)}}{\mathcal{I}_0^{(1)}} \nsp\right) }
{\ds\iint \mathcal{I}_0^{(1)}\!\left(\!1+\sum_{n=1}^3 s_n \frac{\mathcal{I}_n^{(1)}}{\mathcal{I}_0^{(1)}}\nsp\right)\!\ud^2 x}
= w(\bx) \frac{1+\bm{\mu}(\bx)\cdot\bs}{1+\overline{\bm{\mu}}\cdot\bs},
\end{align}
where the abbreviations in Eq.~(\ref{eq:mu}) and the paragraph that follows it were used. 
The elements of the Fisher information matrix may be derived from the log-likelihood function
\begin{equation}
\ell(\bs|\bx) = \ln\!\left[w(\bx)\frac{1+\bm{\mu}(\bx)\cdot\bs}{1+\overline{\bm{\mu}}\cdot{\bs}}\right]\nsp.
\end{equation}
The first two derivatives of this function with respect to the normalized Stokes parameters are 
\begin{align}
\frac{\partial\ell}{\partial s_n} &= \frac{\mu_n}{1+\bm{\mu}\cdot\bs} - \frac{\overline{\mu_n}}{1+\overline{\bm{\mu}}\cdot\bs},
\qquad
\frac{\partial^2\ell}{\partial s_m\partial s_n} 
= -\frac{\mu_m\mu_n}{(1+\bm{\mu}\cdot\bs)^2} + \frac{\overline{\mu_m}\;\overline{\mu_n}}{(1+\overline{\bm{\mu}}\cdot\bs)^2}.
\end{align}
Then the $(m,n)$th element of the unit Fisher information matrix for a single-photon measurement is defined as
\begin{align}
[\mathbb{F}(\bs)]_{mn}
  &= -\iint \left(\frac{\partial^2}{\partial s_m\partial s_n}\ell(\bs|\bx)\right)\!P(\bx|\bs)\sp\ud^2x\nonumber\\[4pt]
  &= -\iint w(\bx) \frac{1+\bm{\mu}\cdot{\bs}}{1+\overline{\bm{\mu}}\cdot\bs}
      \left[ -\frac{\mu_m\mu_n}{(1+\bm{\mu}\cdot\bs)^2} + \frac{\overline{\mu_m}\;\overline{\mu_n}}{(1+\overline{\bm{\mu}}\cdot\bs)^2} \right]\!\ud^2x\nonumber\\[4pt]
  &= \left(\! \frac{1}{1+\overline{\bm{\mu}}\cdot\bs} \nsp\iint\!\frac{\mu_m\mu_n}{1+\bm{\mu}\cdot\bs}w(\bx)\sp\ud^2 x \nsp\right) \!
  - \left(\! \frac{\overline{\mu_m}\;\overline{\mu_n}}{(1+\overline{\bm{\mu}}\cdot\bs)^3} \iint(1+\bm{\mu}\cdot\bs)w(\bx)\sp\ud^2x \nsp\right)\!.
\end{align}
The integral in the second term evaluates to $1+\overline{\bm{\mu}}\cdot\bs$, which cancels with one factor in the denominator. This simplification leads to Eq.~(\ref{eq:Fishergen}). 

\section*{Funding}
National Science Foundation (NSF) (PHY-1507278); Excellence Initiative of Aix Marseille University - A$^*$MIDEX, a French ``Investissements d'Avenir'' programme.

\section*{Acknowledgments}
The authors would like to thank Thomas G. Brown,  Philippe R\'efr\'egier, and Valentine Wasik.

\end{document}